\documentstyle[aps]{revtex}

\small

\begin{document}

\newcommand{\ia}{\'{\i}}



SCES'99 PAPER REFERENCE Nr. : 373.

Version Date: 16 August 1999.

Title: 

{\large SELF-CONSISTENT CLUSTER CPA METHODS AND 

\vspace{0.1cm}

THE NESTED CPA THEORY}

\vspace{0.4cm}

C.I. Ventura$^{a}$, R.A. Barrio$^{b}$.

a- CONICET - Centro At\'omico Bariloche, 8400 Bariloche, ARGENTINA.

b- Inst. de F\ia sica, UNAM, 01000 M\'exico
D.F., MEXICO.

\vspace{0.4cm}

\hspace{1cm}The coherent potential approximation, CPA, is a
useful tool to treat systems with disorder. Cluster theories have been
proposed to go beyond the translation invariant single-site CPA
approximation and include some short range correlations. In this
framework one can also treat simultaneously diagonal disorder 
(in the site-diagonal elements of the Hamiltonian) and non-diagonal
disorder (in the bond energies). 
It proves difficult to obtain reasonable results, free of
non-analyticities, for lattices of dimension higher than one ($D>1$).
 We show electronic structure results obtained for a Hubbard model,
treated in mean field approximation, on a square lattice and a 
simple cubic lattice, with the simultaneous
inclusion of diagonal and non-diagonal disorder. We compare 
the results obtained using three different methods 
to treat the problem: a self-consistent 2-site cluster CPA method, the
Blackman-Esterling-Berk single-site like extension of the CPA   
and a nested CPA approach.  

\vspace{1cm}

Disorder, Coherent Potential Approximation, Cluster Theories.

\vspace{2cm}

{\bf Contact Address:} 

Dr. Cecilia I. Ventura,

Centro At\'omico Bariloche, 

8400 - Bariloche, ARGENTINA.

\vspace{0.5cm}

Tel.: (+54) 2944 - 445 138 / 462 506 

FAX:  (+54) 2944 - 445 299

E-mail: ventura@cab.cnea.gov.ar  

\newpage


Many-site or cluster extensions to the single site CPA introduced by
Soven in 1967 \cite{soven} have been considered
 (see Refs.[2], e.g.) to overcome its limitations. 
Few results are available for $D>1$, while many approaches have 
limited applicability due to spurious 
non-analyticities in the Green's function \cite{nbutler}.   
Analyticity of the theory was proved in a few cases \cite{mh,beb,kaplan}. 
CPA becomes exact in the infinite-D limit \cite{vollhardt}, 
and interesting connections can be made with 
the recently developed methods for strongly correlated systems \cite{infd}. 

We now briefly present our  
electronic structure calculations (a more
complete presentation will be given elsewhere \cite{nuestro}). 
 The Hamiltonian considered here is:
$$H \, =  \, \sum_{j \in C}  E_{C}  c^{+}_{j}c_{j} + 
\sum_{j \in A}  E_{A}  a^{+}_{j}a_{j} 
- t \sum_{<i,j> \in C} c^{+}_{i}c_{j} $$
\begin{equation}
- t  \sum_{<i,j> \in A} 
a^{+}_{i}a_{j} - t'  \sum_{<i,j>, i \in A,j \in C} a^{+}_{i}c_{j} + H.c. 
\end{equation}
$c^{+}_{j} (a^{+}_{j})$ creates an electron in the C(A)-species tight-binding
(TB) orbital centered at site $j$. 
The two components of the alloy, $\alpha= C, A$ respectively, are described
by a single band Hubbard model \cite{hubbard} each, with 
site energies $\varepsilon_{\alpha}$, local correlation
$U_{\alpha}$, and hopping $t$. A mean field treatment for the 
paramagnetic phase of the model maps the problem to a TB  
Hamiltonian with renormalized site energies, here: 
$ E_{C} = \varepsilon_{C} + U_{C} \frac{<n_{c}>}{2} $, and
$ E_{A} = \varepsilon_{A} + U_{A} \frac{<n_{a}>}{2} $. 
$t'$ describes hopping between sites occupied by different species.
 
To treat this model we first employed a  
self-consistent 2-site cluster CPA approach (denoted hereafter 2-CPA) 
\cite{foo,nuestro}. A uniform effective medium is defined, 
through a 2x2 complex self-energy matrix $\Sigma$ 
to be determined self-consistently by the generalized CPA-condition 
for the cluster. 
Concretely, if $H_0$ denotes the tight-binding Hamiltonian of a
D-dimensional lattice with hopping $t$, we define:
\begin{eqnarray}
H_{eff} \, = \, H_{0} \, + \, \Sigma \, \, , \, \, \quad
\Sigma_{i,j} \, = \, (\sigma_{d} - \sigma_{n}) \, \delta_{i,j} +
\sigma_{n}.
\end{eqnarray}
Isolating a cluster composed of two nearest neighbour sites immersed in the
above effective medium, one has the reduced cluster Hamiltonian:
\begin{eqnarray}
H_{(1,2)} \, = \, H_{eff}(\omega) \, + \, \Delta H(\omega), 
\end{eqnarray}   
where $\Delta H(\omega)$ is the 2x2 matrix describing the difference
between the projection of the real Hamiltonian H onto the two sites of the
cluster, and the effective medium. We impose that the average Green
function  should coincide with the
 effective medium Green function, $G_{eff}$. Using Dyson's equation, this ``2-CPA''
condition can be written:
\begin{eqnarray}
< [ I - \Delta H \, G_{eff} ]^{-1} \, \Delta H >_{(1,2)} \, = 0.
\label{2cpa}
\end{eqnarray} 

The second method we employed is a variation of the
2-CPA, the nested CPA approach (NCPA). 
Basically, one can rewrite Eq.~[\ref{2cpa}] in a convenient form 
such that one can split the problem of finding the self-energy
$\Sigma$, in many (five) simpler CPA alloy calculations, which are ``nested'' 
as three consecutive levels of CPA are required. The 
average over the occupations of the 2 sites of the cluster 
implied by  Eq.~[\ref{2cpa}] has the form \cite{nuestro}:
\begin{equation}
P_{cc} \, \chi_{cc} \, + P_{aa} \, \chi_{aa} \, + \, 
P_{ac} \, \chi_{ac} \, + \, P_{ca} \, \chi_{ca} \, = \, 0, 
\end{equation}
where $c$ is the concentration of the A-species, $p$ is a
parameter included to analyze phase segregation, and: $
P_{cc} \, = \, (1-c^2) \, + \, c(1-c)p ,  
P_{aa} \, = \, c^2 \, + \, c(1-c)p ,  P_{ac}=P_{ca}= c(1-c)(1-p) $.
Introducing $\chi_{t} = (\chi_{ac} + \chi_{ca}) / 2 $, it  
can be rearranged in a form suggestive of the
sequence of alloy problems to be solved in the NCPA \cite{nuestro}:
$$
0 \, = \, p \, [ ( 1 - c ) \, \chi_{cc} \, + \, c \, \chi_{aa} ]
 + \, (1-p) \, \left\{ ( 1 - c ) \right. $$
\begin{equation} 
\left. [ ( 1 - c ) \, \chi_{cc} \, + \, c \,
\chi_{t} \, ]  \,
   + \, c \,  [  c \, \chi_{aa} \, + \, ( 1 - c ) \, \chi_{t} ]\right\}.
\end{equation}

Finally, the third method we employed was the single-site extension of
the CPA by Blackman, Esterling and Berk (BEB) \cite{beb} to include
 non-diagonal disorder. 

In Fig. 1 we exhibit the results we obtained with the three methods 
described on a square lattice at c=0.5, for which we used the exact bare 
local and nearest neighbour's Green functions in terms of the complete 
elliptic integral \cite{morita}, continued analytically
onto the physical sheet of the Riemann surface \cite{nuestro}.
 The corrections to
the single site approach of BEB obtained with the 2-CPA and the NCPA are not 
qualitatively very important, except at the band center. In Fig. 2 
we show results for a three dimensional lattice: here we adopted
the semielliptic approximation to the bare Green functions of the
simple cubic lattice.

\newpage

{\bf Figure Captions:}

\vspace{0.3cm}

Fig.1 - Square Lattice: Density of states (DOS) vs.
energy, approximations employed labeled inside plot. Energies in units
of the hopping $t$. Open symbol: C-species local density of states in
2-CPA approx. Parameters: $E_{C}=0$, $E_{A}$=2.5, t'/t=1.1, c=0.5, p=0.    

\vspace{0.3cm}

Fig.2 - Simple Cubic Lattice: DOS vs. energy. 
 $E_{C}=0$, $E_{A}$=2, t'/t=1.2, c=0.25, p=0.   

\end{document}